\documentclass[12pt]{iopart}
\usepackage{iopams}
\usepackage[xdvi]{graphicx}

\newcommand{\bra}{\left\langle}
\newcommand{\ket}{\right\rangle}

\newcommand{\ep}{\epsilon}
\newcommand{\lr}[1]{\left(#1\right)}
\newcommand{\nm}{\nonumber\\}
\newcommand{\phiu}{\phi_{\rm u}}
\newcommand{\tc}{t_{{\rm c}}}

\newcommand{\calF}{{\cal F}}

\newcommand{\gc}{g_{\rm c}}
\newcommand{\fc}{f_{\rm c}}

\newcommand{\Dt}{\Delta t}

\begin{document}


\title
[Singular perturbation near mode-coupling transition]
{Singular perturbation near mode-coupling transition}

\author{Mami Iwata and Shin-ichi Sasa}

\address{Department of Pure and Applied Sciences, University of Tokyo, 
Komaba, Tokyo 153-8902, Japan}

\begin{abstract}
We  study the simplest mode-coupling equation which describes
the time correlation function of the spherical $p$-spin glass 
model. We  formulate a systematic perturbation theory  near 
the mode-coupling transition point by introducing 
multiple time scales. In this formulation,  the invariance 
with respect to the dilatation of time in a late stage yields 
an arbitrary  constant in a leading order expression of the 
solution. The value of this 
constant is determined by a solvability condition associated with 
a  linear singular equation for perturbative corrections in the 
late stage. 
The solution thus constructed provides 
exactly  the $\alpha$-relaxation time. 
\end{abstract}




\section{Introduction}

About a quarter of a century ago, a peculiar type of slow relaxation 
was discovered theoretically in researches of glassy systems 
\cite{Goetze,Leut}. This relaxation behavior is characterized
by two different time scales, both of  which diverge at 
a temperature. 
An illustrative example exhibiting such a behavior 
is the spherical $p$-spin glass model \cite{p-spin}. 
The normalized  time-correlation function $\phi(t)$  
of total magnetization in this model
turned out to satisfy exactly the so-called mode-coupling 
equation, which is written as 
\begin{equation}
\partial_t \phi(t)=-\phi(t)-g \int_0^t ds \phi^2(t-s)\partial_s \phi(s)
\label{MCT}
\end{equation}
for the case $p=3$.  Here, the initial condition is 
given by $\phi(0)=1$, and the parameter $g$ is
proportional to the square of the inverse temperature. Since this equation is 
derived under an  assumption that the system possesses the stationarity, 
(\ref{MCT}) is valid only in a regime $0 \le g < \gc$,
where $\gc$ will be given later. 

A remarkable  feature of (\ref{MCT}) is the existence of nonlinear 
memory. One can  regard (\ref{MCT}) as one of the simplest equations 
that characterize a universality class consisting of models with 
nonlinear memory. Indeed, some qualitatively new features in
glassy systems have been uncovered by studying (\ref{MCT})
and its extended forms. 
(See Ref. \cite{MCT} as a review.) 
In particular, two divergent time scales 
were found just below the  transition point $g = \gc$, 
and  the precise values of the exponents characterizing
the divergences were determined. 
Furthermore, extensive studies have been attempted so as
to construct the solution in a systematic manner. On the
basis of past achievements, in the present paper, we propose 
a perturbation method for analyzing  (\ref{MCT}), which might 
shed  new light on the nature near the mode coupling transition.

We shall address the question we study in this paper.
Let $f_\infty$ be the value of $\phi(t\to \infty)$. 
We substitute $\phi(t)=G(t)+f_\infty$ into (\ref{MCT}) and take 
the limit $t \to \infty$. We then obtain 
\begin{equation}
-f_\infty+g f_\infty^2(1-f_\infty)=0,
\label{f:eq}
\end{equation}
where we have used the relation $G(0)=1-f_\infty$.
From the graph $g f_\infty^2(1-f_\infty)$ as a function of $f_{\infty}$,
we find that the  non-trivial solution ($f_\infty\not =0$) 
appears when $g \ge \gc=4$.  This transition
is called the mode-coupling transition. 
Note that $f_\infty=1/2$ when $g=\gc$. 
Below we express this value of
$f_\infty$ as $\fc$. We then introduce a small positive parameter $\epsilon$
by setting $g=\gc -\epsilon$, and we denote  the solution  of (\ref{MCT}) 
by $\phi_\ep(t)$.  In this paper, we formulate a perturbation 
theory for (\ref{MCT}). As a result, we obtain
an asymptotic form of $\phi_\ep(t)$ in the  small $\epsilon$ limit. 

\begin{figure}[htbp]
\begin{center}
\begin{tabular}{cc}
\includegraphics[width=6cm]{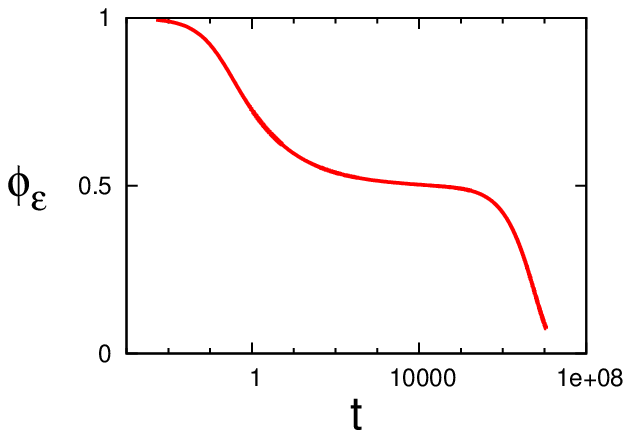}
\includegraphics[width=6cm]{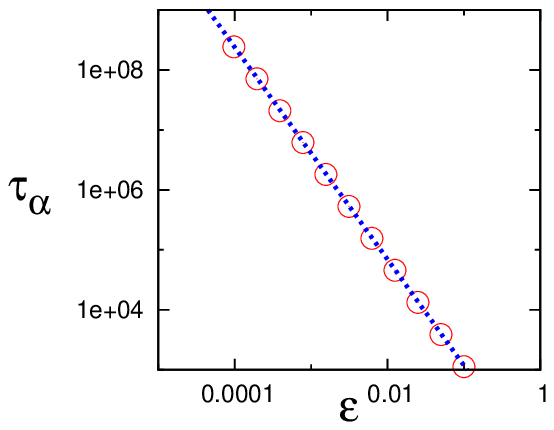}
\end{tabular}
\end{center}
\caption{$\phi_\ep(t)$ with $\ep=10^{-3}$ (left).
$\alpha$-relaxation time as a function of $\epsilon$ (right).
The circle symbols represent the result of numerical simulation of 
(\ref{MCT}). The dotted line corresponds to the theoretical
calculation $\tau_\alpha=20\ep^{-1.77}$ given by (\ref{taualpha}).}
\label{fig4}
\end{figure}

More concretely, for a given small positive $\epsilon$, we want to
express  $\phi_\ep(t)$ in terms of $\ep$ and $\ep$-independent functions. 
For readers' reference, in figure \ref{fig4} (left), we display the numerical 
solution $\phi_\ep(t)$ with $\ep=10^{-3}$. Here, when solving (\ref{MCT}),
we  used the algorithm proposed in Ref. \cite{Fuchs}.  
In figure \ref{fig4} (right), we also display
the $\ep$-dependence of  the $\alpha$-relaxation time $\tau_\alpha$ 
which is defined by  $\phi_\ep(\tau_\alpha)=1/4$.
We want to calculate $\tau_\alpha$ based on  our theory. 

This paper is organized as follows. In section \ref{prelim},
we set up our theory. In particular, we give a useful
expression of a perturbative solution. 
This section also includes a review of known facts in order 
to have  a self-contained description. 
In section \ref{formu}, we formulate a systematic perturbation theory
on the basis of our expression, and we determine a leading order form 
of the solution. We  check the validity of our theory
by comparing our theoretical result of $\tau_\alpha$ with that 
measured by direct numerical simulations  of (\ref{MCT}).
Section \ref{remark} is devoted to remarks on possible
future studies. Technical details are summarized in Appendices.


\section{Preliminaries}\label{prelim}

\subsection{Solution with  $\ep=0$}

\begin{figure}[htbp]
\begin{center}
\begin{tabular}{cc}
\includegraphics[width=6cm]{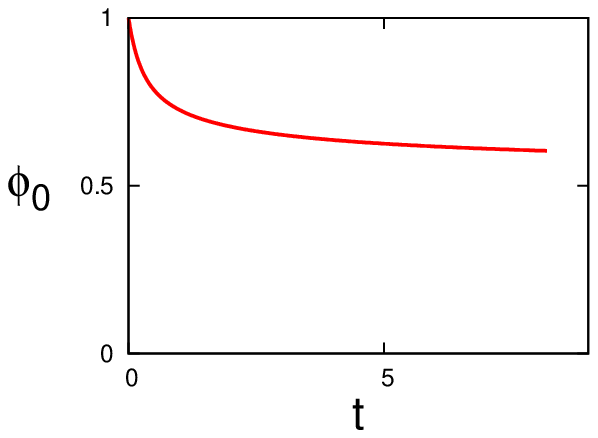}
\includegraphics[width=6cm]{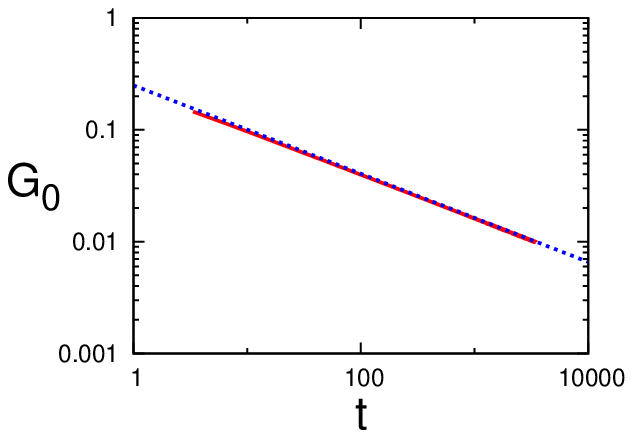}
\end{tabular}
\end{center}
\caption{$\phi_0(t)$ (left) and $G_0(t)$ in the log-log plot (right). 
The dotted line in the right figure represents $0.25t^{-0.395}$.}
\label{fig1}
\end{figure}

We first investigate the solution $\phi_0(t)$.
It is expressed  by $\phi_0(t)=G_0(t)+\fc$ with a function  $G_0(t)$
which decays to $0$ as $ t \to \infty$. The equation for $G_0$ is 
written as
\begin{eqnarray}
\partial_t G_0(t) +\fc+G_0(t) +g_c\int_0^{t}ds 
\lr{\fc +G_0(t-s) }^2 \partial_s G_0(s)=0.
\label{G0t}
\end{eqnarray}
An asymptotic form of $G_0(t)$ in the large $t$ limit can be derived by 
employing a formula
\begin{equation}
x \int_0^t ds[(t-s)^{-x}-t^{-x}]s^{-x-1}
=\left(1-\frac{\Gamma^2(1-x)}{\Gamma(1-2x)} \right)t^{-2x}
\end{equation}
for any $x < 1$ and $t >0$, 
where $\Gamma(x)$ is the Gamma function. The result is 
\begin{eqnarray}
G_0(t) \simeq  c_0 t^{-a},
\label{a:def}
\end{eqnarray}
where $a$ is a  constant that  satisfies a  relation
\begin{equation}
\frac{\Gamma^2(1-a)}{\Gamma(1-2a)}=\frac{1}{2}.
\end{equation}
The value of $a$ is  estimated as $a=0.395$. 
In figure \ref{fig1}, we display the graphs of 
$\phi_0(t)$ and $G_0(t)$,
which are calculated numerically.

As shown in \ref{app:c0},  an approximate expression of $c_0$ is 
calculated as $c_0^{\rm app}=(a/(1-a))^a(1-a)/2^{a+1}$ by a matching 
procedure. Its value,  $0.194$, is not far from $c_0=0.25$ obtained from  
a numerical fitting of the graph of $G_0(t)$. It might be possible 
to improve the approximation in a systematic manner. However,
in this paper, we do not pursue such  improvements. The important 
thing here is that the $\ep$-independent function $G_0(t)$ is defined 
with understanding of its asymptotic form.

\subsection{Expression of the solution with $\ep > 0$}\label{exp:sol}

One may expect that the solution $\phi_\ep(t)$ is close to
$\phi_0(t)$. However, recall that $\phi_0(t \to \infty)$
changes discontinuously from $0$ to $1/2$  when $g$ passes 
at $g=\gc$ from below. This fact means that a  small
perturbation from $g=\gc$ ($\ep=0$) yields a singular 
behavior. Examples of such  {\it singular perturbation} can 
be seen in  Refs. \cite{Holmes, Bender}.


Before formulating effects of the perturbation, we conjecture
a functional form of the solution $\phi_\ep(t)$. First, $\phi_{\ep}(t)$ 
should be close to $\phi_0(t)$ in an  early stage where $\phi_\ep > \fc$.
Since $\phi_0(t) \to \fc$ as $ t \to \infty$, the trajectory $\phi_\ep(t)$
stays in a region near $\fc$ for a long time.  
However, since there is no non-trivial solution $f_\infty \not = 0$ 
for positive $\epsilon$, 
$\phi_\ep$ goes away from the region $\phi_\ep \simeq \fc$
and finally approaches the origin 
$\phi_\ep = 0$. 
Such a behavior is substantially different 
from $\phi_0(t)$. We thus introduce a quantity $A(\ep^{\gamma_2} t)$
that describes the relaxation behavior from $\phi_\ep \simeq \fc$ 
to $\phi_\ep=0$, where $A(0)=\fc$. The functional form of $A$ is 
independent of $\ep$, while its argument is the scaled time 
$t_2=\ep^{\gamma_2} t$ with a positive constant $\gamma_2$. 
We also expect that a switching from $\phi_\ep(t)\simeq \phi_0(t)$ 
in the early stage to $\phi_\ep(t)\simeq A(\ep^{\gamma_2}t)$ in 
the late stage occurs around another characteristic time of 
$O( \ep^{-\gamma_1})$ with a positive constant $\gamma_1$. 
Keeping this behavior in mind, we  express the solution  as
\begin{eqnarray}
\phi_{\ep}(t) = G_0(t) \Theta(\ep^{\gamma_1}t)
+ A(\ep^{\gamma_2}t)+ \varphi_\ep(t), 
\label{rep}
\end{eqnarray}
where the switching function $\Theta$ satisfies $\Theta(0)=1$ and 
$\Theta(\infty)=0$. 
The functional form of 
$\Theta$ is independent of $\ep$, while its argument depends on $\ep$. 
$\varphi_\ep(t)$ represents a small correction 
that satisfies $\varphi_\ep \to 0$ in the limit 
$\ep \to 0$ for any $t$. $\Theta$, $A$, $\gamma_1$,  $\gamma_2$ and 
$\varphi_\ep(t)$ will be determined later. 


In order to have a simple description, 
we define a set of scaled coordinates  
$(t_0, t_1, t_2)$ on the time axis as $t_i=\ep^{\gamma_i} t$, 
where $\gamma_0=0$. Throughout the paper, a time coordinate
with an integer subscript represents the scaled coordinate 
determined by the subscript. Note that  $t_0$, $t_1$  
and  $t_2$ appear as the arguments of $G_0$, $\Theta$ and $A$, 
respectively. Physically, the first relaxation occurs in the early stage 
$t_0 \sim O(1)$; the behavior around $\phi_\ep=\fc$ is observed in the 
intermediate stage $t_1 \sim O(1)$; and the relaxation behavior from 
$\phi_\ep \simeq \fc$ to $\phi_\ep=0$ is described in the late stage 
$t_2 \sim O(1)$.  In researches of glassy systems, 
the intermediate
and the late stages are termed  the $\beta$-relaxation regime
and the $\alpha$-relaxation regime, respectively. 

\subsection{Equation for A} 

We substitute (\ref{rep}) into (\ref{MCT}) and take 
the limit $\ep \to 0$ with the scaled time $t_2=\ep^{\gamma_2} t$ fixed.
We then obtain
\begin{eqnarray}
{A}(t_2) -g_c(1-\fc) {A}^2(t_2)
+ g_c\int_{0}^{t_2}ds_2 {A}^2(t_2-s_2) {A}'(s_2)=0.
\label{def_A}
\end{eqnarray}
In this paper, the prime symbol represents the differentiation
with respect to the argument of the function. 
The equation (\ref{def_A}) 
provides the explicit definition of $A$ with the condition 
$A(0)=\fc$.  However, this equation 
{\it cannot} determine $A$ uniquely.
Indeed, for a given solution  $ A(t_2)$  of  (\ref{def_A}), 
$ A(\lambda t_2)$ with  any positive $\lambda$ is another 
solution of (\ref{def_A}). This {\it dilatational 
symmetry} is a remarkable property of (\ref{def_A}). 

\begin{figure}[htbp]
\begin{center}
\begin{tabular}{cc}
\includegraphics[width=6cm]{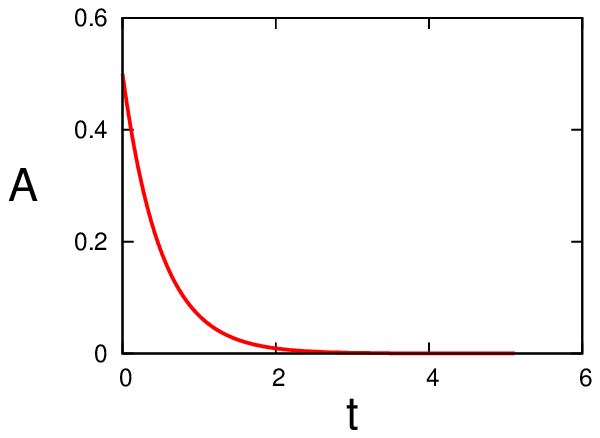}
\includegraphics[width=6cm]{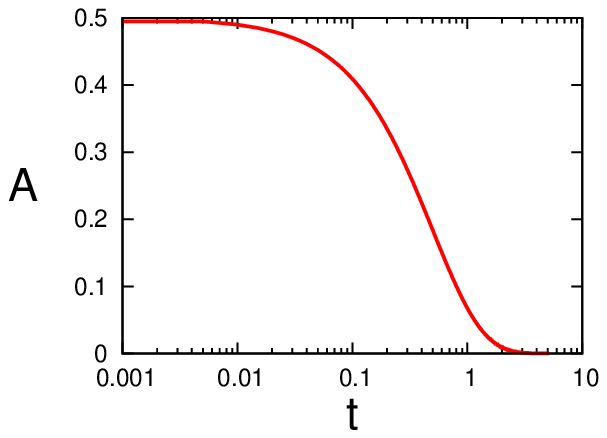}
\end{tabular}
\end{center}
\caption{$A(t)$ (left) and its semi-log plot (right). }
\label{fig2}
\end{figure}

Here, by analyzing the short time behavior in 
(\ref{def_A}), one can confirm that $A(t_2)-\fc$ 
is proportional to $t_2$ in the small $t_2$ limit.
Thus, we  can choose a  special solution of
$A$ such that  $A'(0)=-1$. In the argument below,  $A$ represents this
special solution; and the other solutions are described by
$A(\lambda t_2)$. For later convenience, we  define $A_\lambda$
by $ A_\lambda(t_2)=A(\lambda t_2)$, and  $A$ in (\ref{rep}) 
is replaced with $A_\lambda$. 
In particular, we have
\begin{equation}
A_\lambda(t_2)=\fc-\lambda t_2 +o(t_2)
\label{A:scale}
\end{equation}
in the limit $t_2 \to 0$. Note that $\lambda$ is an arbitrary constant
until a special requirement is imposed. 
The functional form of $A(t_2)$ can be obtained by solving  (\ref{def_A}) 
numerically with  a simple discretization of time. 
We display the graph of $A(t_2)$ in figure \ref{fig2}. 
It should be noted here that the mathematical determination of the 
functional form is not the heart of
the problem. The important thing is that the  $\ep$-independent function 
$A$ is defined without any ambiguities. 

We express the dilatational symmetry in terms of a mathematical 
equality. Let us define 
\begin{equation}
\Phi_2(t_2;A_\lambda)\equiv
A_\lambda(t_2) -\gc(1-\fc)A_\lambda^2(t_2)
+ \gc\int_{0}^{t_2}ds_2  A_\lambda^2(t_2-s_2)  A_\lambda'(s_2).
\label{def_A_2}
\end{equation}
Since  (\ref{def_A}) is identical to $\Phi_2(t_2;A)=0$,
the dilatational symmetry  is expressed by 
$\Phi(t_2; A_\lambda )=0$ for any $\lambda$.
Then, taking the derivative with respect to $\lambda$, 
we obtain 
\begin{eqnarray}
\int_0^\infty ds_2 L_A(t_2,s_2) \partial_{\lambda} A_\lambda(s_2)=0,
\label{L_zero}
\end{eqnarray}
where $L_A$ is the linearized operator around $A_\lambda$, 
which is defined by 
\begin{equation}
L_A(t_2,s_2)=\frac{\delta \Phi_2(t_2;A_\lambda)}{\delta A_\lambda(s_2)}.
\end{equation}
Its  explicit form is given by 
\begin{eqnarray}
L_A(t_2,s_2)
&=&
\delta(t_2-s_2)
\lr{1-2\gc(1-\fc) A_\lambda(s_2)+\gc \fc^2}\nm
&&+\gc\theta(t_2-s_2)2
A_\lambda'(t_2-s_2)
\lr{
A_\lambda(s_2)+ A_\lambda(t_2-s_2)}.
\label{cn_def}
\end{eqnarray}
It will be found below that  (\ref{L_zero})  plays a key role in our
formulation.

Now, from the definition of $\gamma_1$, we have 
$\fc+G_0(\ep^{-\gamma_1})\simeq A_\lambda(\ep^{\gamma_2-\gamma_1})$.
By substituting  (\ref{a:def}) and (\ref{A:scale})
into this relation, we obtain
\begin{equation}
\gamma_1 =\frac{\gamma_2}{1+a}.
\label{gamma1:det1}
\end{equation}

\subsection{Functional form of $\Theta$}

We can formulate a systematic perturbation theory 
with employing an arbitrary switching function $\Theta(t_1)$ 
when  it decays faster than a power-law function 
$t_1^{-1+a}$, as we will see in the next section. For  example, 
one can choose a physically reasonable form
\begin{equation}
\Theta(t_1)=\exp(-t_1/\tc),
\label{theta:ex}
\end{equation}
where $t=\tc \ep^{-\gamma_1}$ corresponds to the time when the graph
$\phi_0(t)$ is closest to that of $A_\lambda (\ep^{\gamma_2} t)$. 
That is, $\tc$ satisfies $a c_0 \tc^{-a-1}=\lambda $. 
Note, however, that there is no reason that we must choose
this form. Indeed,  other forms such as $ \Theta(t_1)=\exp(-2t_1/\tc)$ 
and  $ \Theta(t_1)=\exp(-(t_1/\tc)^2)$  might also be physically
reasonable to the same extent as (\ref{theta:ex}).
Of course, the final result should be independent 
of the choice of the functional form. 

\subsection{Summary}

In our formulation, we set the unperturbative solution $\phiu$ as
\begin{eqnarray}
\phiu(t) = G_0(t) \Theta(\ep^{\gamma_1}t)
+ A_\lambda (\ep^{\gamma_2}t), 
\label{unpsol}
\end{eqnarray}
and we  express the perturbative solution $\phi_\ep$ by
\begin{equation}
\phi_\ep(t)=\phiu(t)+\varphi_\ep(t).
\label{sol:rep:f}
\end{equation}
$G_0$ and $A$ are already determined.  $\gamma_1$ is connected
with $\gamma_2$ in (\ref{gamma1:det1}). $\Theta$ is assumed to take 
an arbitrary form. Thus, the problem we  solve is the determination of 
$\gamma_2$ and $\lambda$ as well as the perturbative calculation of 
the correction $\varphi_\ep(t)$. Note that $\gamma_2$ and $\lambda$ 
appear in the leading order expression of the solution. In particular,
since the value of $\lambda$ has never been known, 
the calculation of $\lambda$ is a cornerstone of our theory.

\section{Systematic perturbation} \label{formu}

\subsection{preliminary}

For any trajectory $\psi(t)$ with $\psi(0)=1$, we define 
\begin{equation}
F_\ep(t;\psi) \equiv  
\partial_t \psi +\psi +g\int_0^{t}ds \psi^2(t-s) \partial_s \psi(s).
\label{Fdef}
\end{equation}
Let  $L_\ep(t,s;\psi)$ be the linearized operator of $ F_\ep(t;\psi)$,
which is defined by
\begin{eqnarray}
L_\ep(t,s;\psi)=\frac{\delta F_\ep(t;\psi)}{\delta \psi(s)}.
\label{phi_model1}
\end{eqnarray}
Its explicit form is  written as 
\begin{eqnarray}
L_\ep(t,s;\psi)&=&\delta(t-s)(1+g)+\delta'(t-s) \nm
& & -g\theta(t-s)\lr{\partial_s\psi^2(t-s)-2 \psi(s)\psi'(t-s)}.
\label{Ldef}
\end{eqnarray}
The mode-coupling equation (\ref{MCT}) is expressed by 
\begin{eqnarray}
F_\ep(t;\phi_\ep)&=&0.
\label{phi_model0}
\end{eqnarray}

\subsection{Calculation}\label{calc}

The substitution of (\ref{sol:rep:f}) into (\ref{phi_model0})
yields non-trivial $\ep$ dependences through the evaluation 
of the integration term in (\ref{Fdef}) with using the 
scaled coordinates. In order to avoid a complicated description, 
we focus our presentation on an important part of the calculation.

We first evaluate $F_\ep(\ep^{-\gamma_2} t_2;\phiu)$ in 
the small $\epsilon$ limit with  $t_2$ fixed.
As explained in \ref{app}, we derive
\begin{equation}
F_\ep(\ep^{-\gamma_2} t_2;\phiu)
\simeq  \ep \frac{1}{\gc}A_\lambda(t_2)
+\ep^{\gamma_2-\gamma_1(1-a)}
2\gc  (\fc+A_\lambda(t_2))A_\lambda^\prime(t_2) 
c_0 \theta,
\label{F_ep}
\end{equation}
where higher order terms of $O(\ep^{\gamma_2-\gamma_1(1-2a)})$
are neglected, and $\theta$ is a constant determined by
\begin{equation}
\theta=\int_0^\infty ds s^{-a}\Theta(s).
\label{c1_def}
\end{equation}
We here make three important remarks on  (\ref{F_ep}). 
First, if we did  not introduce the switching function $\Theta$ 
in the expression of the solution (\ref{rep}), we would have
a form rather different from (\ref{F_ep}), for  which 
the analysis seems to be hard.
Second, the function $\Theta$ should provide a  finite value
of $\theta$. This means that $\Theta(t_1)$ 
should decay faster than  a power-law function $t_1^{-1+a}$. 
Third, the two terms in (\ref{F_ep}) should balance each 
other. 
Otherwise, a contradiction  occurs. (See an argument below
(\ref{final}).) The last remark leads to the relation 
$\gamma_2-\gamma_1(1-a)=1$. By combining it  
with (\ref{gamma1:det1}), 
we obtain well-known results
\begin{equation}
\gamma_1=\frac{1}{2a},
\label{g1}
\end{equation}
and
\begin{equation}
\gamma_2=\frac{1}{2a}+\frac{1}{2},
\label{g2}
\end{equation}
which correspond to the exponents characterizing 
divergences of the $\beta$-relaxation time  and the 
$\alpha$-relaxation time in glassy systems, respectively. 
Then,  (\ref{F_ep}) with (\ref{g1}) and (\ref{g2}) becomes
\begin{equation}
F_\ep(\ep^{-\gamma_2}t_2;\phiu)=
\ep\calF_2^{(1)}(t_2)+O(\ep^{3/2}),
\label{Fexp}
\end{equation}
where $\calF_2^{(1)}$ is the  $\ep$-independent 
function given by 
\begin{equation}
\calF_2^{(1)}(t_2) 
= \frac{1}{\gc}A_\lambda(t_2)
+2\gc  (\fc+A_\lambda (t_2))A_\lambda^\prime( t_2) c_0 \theta.
\label{F_eporder}
\end{equation}
Furthermore, we  can prove
\begin{equation}
\ep^{-\gamma_2} L_\ep(\ep^{-\gamma_2}t_2,\ep^{-\gamma_2}s_2;\phiu)
=L_A(t_2,s_2)+O(\ep),
\label{Lexp}
\end{equation}
where $L_A$ is given by (\ref{cn_def}). 

Now, let us compare  (\ref{Fexp}) and  (\ref{phi_model0})
with (\ref{sol:rep:f}). We then find that the perturbative correction 
is expressed as 
\begin{equation}
\varphi_\ep( \ep^{-\gamma_2}t_2)  
=\ep \bar \varphi_2^{(1)}(t_2)+ O(\ep^{3/2})
\label{sol:exp}
\end{equation}
in the regime $t_2 \sim O(1)$ with the limit $\ep \to 0$. 
We also write
\begin{equation}
\varphi_\ep( \ep^{-\gamma_1}t_1)  =\ep^{\alpha}\bar \varphi_1^{(\alpha)}(t_1)
+o(\ep^{\alpha})
\label{alpha-def}
\end{equation}
in the regime $t_1 \sim O(1)$ with the limit $\ep \to 0$,
where  $\alpha$ is a positive constant. 
Then, (\ref{phi_model0}) with (\ref{sol:rep:f})
becomes 
\begin{equation}
F_{\ep}(\ep^{-\gamma_2} t_2;\phiu+\ep^{\alpha}\bar \varphi_1^{(\alpha)})
+\ep
\int_0^\infty  ds_2 L_A(t_2,s_2) \bar \varphi_2^{(1)}(s_2)
=o(\ep),
\end{equation}
where the contribution of $\varphi_\ep(t_0)$ is included in
the right-hand side. 
Here, by an argument similar to
\ref{app}, we can estimate 
\begin{equation}
F_{\ep}(\ep^{-\gamma_2} t_2;\phiu+\ep^{\alpha}\bar \varphi_1^{(\alpha)})
=\ep \calF_2^{(1)}(t_2)+O(\ep^{\alpha+1/2}).
\label{estF}
\end{equation}
In order to describe a theoretical framework in a simple manner,
for the moment, we focus on the case $\alpha >1/2$. The other 
case $\alpha  \le 1/2$  will be discussed  in section \ref{theta:sec}.
The equation for $\bar \varphi_2^{(1)}(t_2)$ is then simply
written as 
\begin{equation}
\int_0^\infty  ds_2 L_A(t_2,s_2) \bar \varphi_2^{(1)}(s_2)
= - \calF_{2}^{(1)}(t_2).
\label{eq1}
\end{equation}

\subsection{Solvability condition}\label{solv:sec}

We notice that (\ref{eq1}) is  a  
linear equation for $\bar \varphi_2^{(1)}$,
which is singular because there exists the zero eigenvector 
$\Phi_0=\partial_\lambda A_\lambda$ associated 
with  the dilatational symmetry (\ref{L_zero}).
Let $\Phi_0^\dagger$ be the adjoint zero eigenvector that satisfies
\begin{equation}
\int_0^\infty  ds_2 L_{A}(s_2,t_2) \Phi_0^\dagger(s_2)=0.
\label{adj}
\end{equation}
Then, there exists a solution of (\ref{eq1}) only when the condition 
\begin{equation}
\int_0^\infty dt_2 \Phi_0^\dagger(t_2) \calF_{2}^{(1)}(t_2)=0
\label{solv}
\end{equation}
is satisfied. Otherwise,  (\ref{eq1}) leads to $0 \not = 0$
and hence there is no solution $\bar \varphi_2^{(1)}$.
The equality (\ref{solv}) is called the {\it solvability condition}
for the singular equation (\ref{eq1}). Note,  however, that 
the solvability condition is not satisfied as an identity. 
Here, let us recall that $\lambda$ is  still an arbitrary
constant. Thus, we are allowed to determine the value of $\lambda$ 
so that the solvability condition (\ref{solv}) can be satisfied.
Only for this special value of $\lambda$, the perturbation theory
can be formulated consistently.

Concretely, 
since we find that $\Phi_0^\dagger(s_2)=\delta(s_2)$
from (\ref{adj}) with (\ref{cn_def}), 
the solvability condition (\ref{solv}) becomes 
\begin{equation}
\calF_{2}^{(1)}(0)=0.
\label{sol-conc}
\end{equation}
The explicit form $\calF_{2}^{(1)}(0)$ obtained from  (\ref{F_eporder}) 
leads to 
\begin{equation}
\lambda=\frac{1}{64c_0 \theta}.
\label{final}
\end{equation}
Here, let us go back to (\ref{F_ep}). 
If $\gamma_2-\gamma_1(1-a)$ were not equal to 1, the condition
(\ref{sol-conc}) could not be satisfied for any positive $\lambda$. 
In this sense, one can regard that the solvability condition determines
the exponent $\gamma_2$ as well as the constant $\lambda$.

\subsection{Determination of $\lambda$}\label{theta:sec}

Apparently, (\ref{final}) shows that 
$\lambda$ depends on the choice of $\Theta$. However, the situation
is a little bit complicated. We shall explain the way how to 
determine the value of $\lambda$ in detail. 

We study the case  $\ep \to 0$ with $t_1$ fixed.
In this limit,  (\ref{rep}) can be expressed as
\begin{equation}
\phi_{\ep}(\ep^{-\gamma_1}t_1) 
=\fc +\ep^{1/2} [c_0 t_1^{-a} \Theta(t_1)-\lambda t_1]
+\varphi_\ep(\ep^{-\gamma_1} t_1),
\label{rep-I}
\end{equation}
where we have used (\ref{a:def}) and (\ref{A:scale}).  Since $\Theta$
is arbitrary, $\varphi_\ep(\ep^{-\gamma_1} t_1)$ includes 
a term of $O(\ep^{1/2})$ unless a special $\Theta$  is employed.
This means that $\alpha$ in (\ref{alpha-def}) is equal to 1/2. 
As is seen from (\ref{estF}), when $\alpha=1/2$, 
$\bar \varphi_2^{(1)}(t_2)$ must be calculated 
with taking account of $\bar \varphi_1^{(1/2)}$. Concretely,
the right hand side of (\ref{eq1}) should be replaced with 
$- \tilde \calF_{2}^{(1)}(t_2)$, where 
\begin{equation}
F_{\ep}(\ep^{-\gamma_2} t_2;\phiu+\ep^{1/2}\bar \varphi_1^{(1/2)})
=\ep \tilde \calF_2^{(1)}(t_2)+o(\ep).
\end{equation}
Then, the solvability condition (\ref{sol-conc}) is also 
replaced with $\tilde \calF_{2}^{(1)}(0)=0$. 
Without the replacement,  
(\ref{final}) provides nothing more than an approximation 
of $\lambda$. For example, (\ref{final}) with (\ref{theta:ex})
leads to $\lambda=[(64\Gamma(1-a))^{(1+a)/(2a)}
c_0^{1/a} a^{(1-a)/(2a)}]^{-1} \simeq 0.022$ as one approximate value.

Now, let us calculate the precise value of $\lambda$.
One natural method is to choose a functional form of $\Theta$ 
so that the condition  $\alpha >1/2$ is satisfied. 
We denote this special $\Theta$ by $\Theta_*$. 
Then, for a given $\Theta$, the 
correction $\bar \varphi_1^{(1/2)}$ is determined by 
\begin{equation}
c_0 t_1^{-a} \Theta(t_1)+\bar\varphi_1^{(1/2)}(t_1)
=c_0 t_1^{-a}\Theta_*(t_1).
\end{equation}
Therefore, (\ref{sol-conc}) using $\Theta_*$ is equivalent to
$\tilde \calF_{2}^{(1)}(0)=0$ using $\Theta$. In other words,
the precise calculation of $\lambda$ starting from $\Theta$
can be done through $\Theta_*$. This also indicates 
explicitly that the final and precise result is independent
of the choice of $\Theta$. In any cases, the problem is focused
on the calculation of  $\Theta_*$.

As explained in \ref{app:eqQ}, 
we can derive the equation for 
$Q(t_1)\equiv c_0 t_1^{-a} \Theta_*(t_1)$ in the form
\begin{eqnarray}
\frac{1}{8}-8\lambda \int_0^{t_1} ds_1 [Q(s_1)-Q(t_1)/2]
\nonumber \\
+2 Q^2+4 \int_{0}^{t_1}ds_1 [Q(t_1-s_1)-Q(t_1)] Q'(s_1)=0.
\label{Q:det}
\end{eqnarray}
We study this equation by regarding $\lambda$ as a parameter
whose value is not specified beforehand. We denote this solution by 
$Q(t_1;\lambda)$. For  almost all $\lambda$,  $Q(t_1;\lambda)$ 
is not bounded as $t_1 \to \infty$, while there exists the special 
value $\lambda_*$
such that $Q(t_1;\lambda_*) \to 0$ as $t_1 \to \infty$. 
A necessary condition for this property 
is easily derived by considering the limit 
$t_1 \to \infty$ in (\ref{Q:det}):
\begin{equation}
\lambda_*=\frac{1}{64}
\left[\int_0^{\infty} ds_1 Q(s_1;\lambda_*) \right]^{-1}.
\label{scon}
\end{equation}
This is equivalent to the expression  (\ref{final}) that determines
the value of $\lambda$ by the solvability condition 
(\ref{sol-conc}) under the assumption $\alpha >1/2$.
Therefore, once we find $\lambda_*$ such that $Q(t_1;\lambda_*)
\to 0$ as $t_1 \to \infty$, this $\lambda_*$ is the precise value
of $\lambda$ that we want to have. Simultaneously, we obtain 
$\Theta_*(t_1)$ from $Q(t_1;\lambda_*)$. 

\begin{figure}[htbp]
\begin{center}
\begin{tabular}{cc}
\includegraphics[width=6cm]{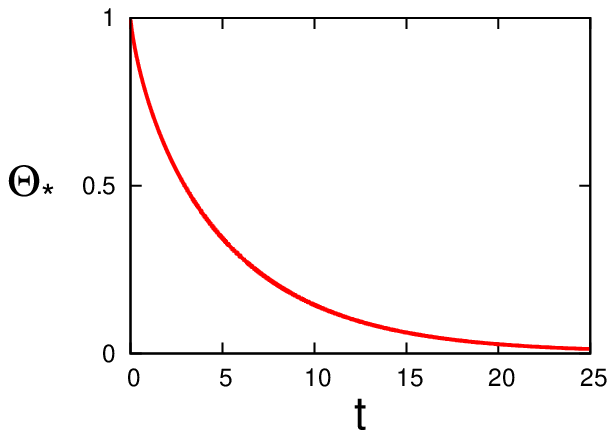}
\includegraphics[width=6cm]{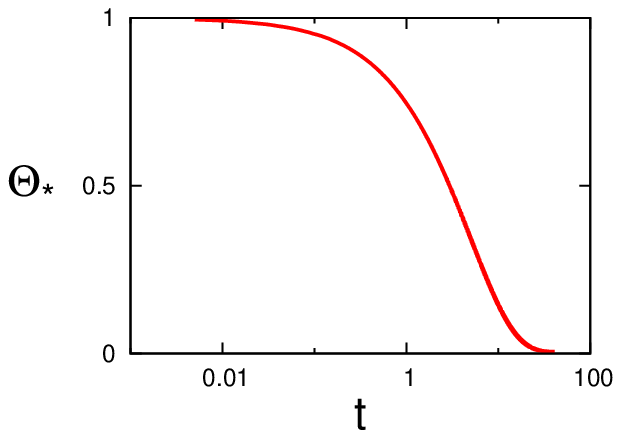}
\end{tabular}
\end{center}
\caption{$\Theta_*(t)$ (left) and its semi-log plot (right). 
}
\label{fig3}
\end{figure}

The problem of finding $\lambda_*$ 
is investigated by a shooting method. We first solve (\ref{Q:det}) 
numerically for a given $\lambda$. Basically, we employ 
a  simple discretization method. 
In order to treat properly the singular behavior 
near $t=0$, we utilize  the result of the short time expansion 
of $Q(t_1;\lambda)$ near $t=0$. 
(See \ref{app:exp} for the short time expansion.)  
Suppose that we already investigated the system with $\lambda_k$, $k=0,1,
\cdots, n$. We here note that $Q(t_1;\lambda) \to -\infty$ as 
$t_1 \to \infty$ when  $\lambda=0$ and that  $Q(t_1;\lambda) \to \infty$ as 
$t_1 \to \infty$ when  $\lambda$ is sufficiently large. Based on  this
observation, we define $\underline{\mu}_n\equiv \max \lambda_k$ such 
that $Q(t_1;\lambda_k) 
\to -\infty$ as $t_1 \to \infty$,  and $\overline{\mu}_n\equiv\min 
\lambda_k$ such that $Q(t_1;\lambda_k) \to \infty$ as $t_1 \to \infty$.
We then  choose $\lambda_{n+1}$ as $\lambda_{n+1}=
(\underline{\mu}_n+\overline{\mu}_n)/2$. Starting from $\lambda_0=0$
and $\lambda_1=1$, we can determine the sequence $\{\lambda_n\}$ for
which $\lambda_\infty=\lim_{n \to \infty}\lambda_n$ exists. 
From the construction method, $Q(t_1;\lambda_\infty) \to 0$ 
as $t_1 \to \infty$. Therefore, $\lambda_*$ is given by 
$\lambda_\infty$.
By performing this procedure numerically,  
we estimate  $\lambda_*=0.017$. 
In this manner, we have determined the precise value of $\lambda$
and the function $\Theta_*$. 
We display the functional form  of $\Theta_*$ in figure \ref{fig3}. 

\subsection{Remarks}\label{result}

At the end of this section, we make two remarks. 
First, as a demonstration of our result, we study the $\alpha$-relaxation 
time $\tau_\alpha$ defined by $\phi_\ep(\tau_\alpha)=1/4$. 
Let $\tau_A$ be $A(\tau_A)=1/4$. Then, 
from the expression of the solution (\ref{sol:rep:f}),
$\tau_\alpha$ is estimated as $\tau_\alpha=(\tau_A/\lambda) \ep^{-\gamma_2}$.
By using the value $\tau_A=0.346$ obtained from numerical integration 
of (\ref{def_A}), we arrive at the theoretical prediction
\begin{equation}
\tau_\alpha=20\ep^{-1.77}.
\label{taualpha}
\end{equation}
In figure \ref{fig4} (right),  
we display the result of numerical simulations
of (\ref{MCT}) with $\ep=0.1\times 2^{-j}$, $j=0,\cdots, 10$. 
The numerical data are perfectly 
placed on the theoretical result (\ref{taualpha}). 
This is  an
evidence that the expression (\ref{final}) is correct.

The second remark is on the systematic formulation. 
In principle, higher order terms such as $\bar \varphi_j^{(3/2)}(t_j)$ 
and $\bar \varphi_j^{(\gamma_2)}(t_j)$ can also be calculated
in a manner similar to that described in sections 
\ref{calc} and \ref{solv:sec}. Such a perturbation theory with 
using a solvability condition has been employed 
in many problems \cite{Bogoliubov, Kuramoto, Cross}. 


\section{Concluding remarks}\label{remark}

We have formulated a systematic perturbation theory
for (\ref{MCT}). Due to the dilatational symmetry (\ref{L_zero}), 
an arbitrary constant $\lambda$ appears in the unperturbed solution 
$\phiu(t)$. Then, the  value of $\lambda$ is determined by 
the solvability condition (\ref{solv}) associated with the linear 
equation (\ref{eq1}) for the perturbative correction  $\bar \varphi_2^{(1)}$. 
The advantage of our systematic perturbation is in a possibility of 
developing new and important directions. 
Concretely, following the three problems will be studied soon.

The first problem is to derive the  fluctuation intensity 
of $\hat C(t,t')=\sum_{jk} \sigma_j(t)\sigma_k(t')/N$ 
just below the mode-coupling transition point for the spherical $p$-spin
glass, where $\sigma_j$ is a real spin variable that satisfies the
spherical constraint $\sum_{j=1}^N \sigma_j^2=1$. 
Note that $\phi(t-t')= C(t-t')/C(0)$ with 
$C(t-t')=\bra \hat C(t,t') \ket $ in the stationary regime.
In a  straightforward approach, one may study an  effective 
potential for $\hat C(t,t')$ \cite{CJT}. Indeed, by employing
a diagrammatic expansion with neglecting vertex corrections, 
the singular behavior of the effective potential was evaluated \cite{BB}. 
Then, since the minimum of the potential corresponds to the 
solution of (\ref{MCT}), the existence of the dilatational 
symmetry yields the Goldstone 
mode which carries a divergent part of fluctuations in the late 
stage. More explicitly, $\lambda$ in our expression is treated 
as a fluctuating quantity, and it is identified with the Goldstone 
mode.  (A Related description of  fluctuations near another
type of bifurcation points can be seen in Refs. 
\cite{Iwata1,Iwata2}.)  The analysis
along this line will shed a new light to the understanding
of fluctuations near mode-coupling transition points. 

As an alternative approach to the description of 
fluctuations near the mode-coupling transition, response properties 
against an auxiliary external field conjugated to $\hat C$ \cite{Franz} and 
against a  one-body potential field \cite{Miyazaki} were investigated.
Their studies successfully derived the scaling form of a singular part
of the fluctuation intensity of $\hat C$ based on an idea that such response 
functions are  related to the fluctuation intensity.
The extension of our work so as to describe the responses  may provide 
a more quantitative result than the scaling form. 
Such an extension 
is related to a study of dynamical behavior in the aging regime,
because its behavior is described by a coupled equation of the 
time correlation function  $C(t,t')$ and 
the response function $R(t,t')$, which are functions of two times
 \cite{Culiandolo}. In addition to a
complicated structure of the equation, the dilatational symmetry
is replaced with the time reparameterization symmetry. Since the 
symmetry is much wider than the dilatational symmetry, 
several new features may appear in the analysis. 
See Ref. \cite{Culiandolo2} as a review for the argument on the 
basis of the time reparameterization symmetry.

The third problem is to analyze a rather wide class
of systems with nonlinear memory.  The qualitative 
change of the solution $f_\infty$ of (\ref{f:eq})
is the same type as that observed in an elementary
saddle-node bifurcation \cite{Guckenheimer}. 
Despite this similarity, the dynamical behavior near the saddle-node
bifurcation is much simpler than that of (\ref{MCT})
owing to the lack of nonlinear memory.
Note that an edge deletion process of $k$-core percolation 
in a random graph is precisely described by a saddle-node bifurcation
\cite{kcore} and it has been pointed out  that  $k$-core
percolation problems are related to jamming transitions \cite{Silbert}. 
Since nonlinear memory effects might appear in jamming transitions,
it is important to study a mixed type of dynamical systems
which connect the elementary saddle-node bifurcation
with the mode-coupling transition. 
The calculation presented in this paper may be useful
in the analysis of such  models.

By studying these problems, we will have deeper understanding of 
slow relaxation with nonlinear memory.  We also hope that 
this paper provokes mathematical studies of the simplest
mode-coupling equation (\ref{MCT}).

\ack

This work was supported by a grant from the Ministry of 
Education, Science, Sports and Culture of Japan, 
No. 19540394. Mami Iwata acknowledges the support  by Hayashi 
memorial foundation for female natural scientists.

\appendix

\section{Approximate expression  of $c_0$}\label{app:c0}

We perform  a short time expansion 
\begin{equation}
G_0(t)=\sum_{n=0}^\infty g_n t^n,
\label{short}
\end{equation}
which is valid around $t=0$. All the coefficients $g_n$ 
can be determined from
a recursive formula. Concretely, $g_0=1/2$, $g_1=-1$ and $g_2=5/2$. 
Respecting the lowest order result $G_0(t)=g_0+g_1t$, we assume 
\begin{equation}
G_0(t)=\frac{1}{2(1+2t)}
\end{equation}
for $t \le t_*$, where $t_*$ will be determined later.
On the other hand, from the asymptotic form 
\begin{equation}
G_0(t)=c_0 t^{-a}
\label{large}
\end{equation}
in the limit $ t \to \infty$, we assume
$G_0(t)=c_0 t_*^{-a}$ for $t \ge t_*$.
Since $G_0(t)$ is smooth at $t=t_*$, we have 
\begin{eqnarray}
\frac{1}{1+2t_*} &=& 2c_0 t_*^{-a}, \\
\frac{2}{(1+2t_*)^2} &=& 2a c_0 t_*^{-a-1}.
\end{eqnarray}
These equations lead to $t_*=a/[(1-a)2]$ and
$c_0=(a/(1-a))^a(1-a)/2^{a+1}$.


\section{Derivation of $\calF_2^{(1)}$} \label{app}

We shall extract a leading order contribution of
$F_2(\ep^{-\gamma_2}t_2;\phiu)$ in the limit $\ep \to 0$.
In order to simplify the calculation, 
we utilize  an identity 
\begin{eqnarray}
\int_0^t ds f(t-s)g'(s)
= &&\int_{t/2}^t ds [f(t-s)g'(s)+g(t-s)f'(s)] \nonumber \\
 && -f(t)g(0)+f(t/2)g(t/2),
\end{eqnarray}
which plays a key role  in an efficient numerical integration
algorithm for solving  mode-coupling equations \cite{Fuchs}. 
By substituting (\ref{unpsol}) into (\ref{Fdef}) and using 
this identity, we  obtain
\begin{eqnarray}
&& F_\ep(\ep^{-\gamma_2} t_2;\phiu) 
=A_\lambda'(t_2)+A_\lambda( t_2)
-g[A_\lambda^2( t_2)-A_\lambda^3( t_2/2)] \nonumber \\
&& 
+g \int_{t_2/2}^{t_2} ds_2 
A_\lambda^\prime( s_2)\left[\phiu^2(\ep^{-\gamma_2}(t_2-s_2))
+2A_\lambda( s_2)\phiu(\ep^{-\gamma_2}(t_2-s_2))
\right].
\label{Fstart}
\end{eqnarray}
Here, we take $\Delta t$ satisfying $\ep^{\gamma_2-\gamma_1} 
\ll  \Delta t \ll 1  $ such that 
$\phiu(\ep^{-\gamma_2} t_2) \simeq A_\lambda (t_2)$ 
in the regime $ \Delta t \le t_2 \le \infty$. 
More explicitly, we assume  
$\Delta t=\ep^{\gamma'}$ with $\gamma_2-\gamma_1 > \gamma' >0$.
We  divide the integration  regime in  the second line 
of (\ref{Fstart}) into two parts, $[t_2/2, t_2-\Delta t]$ and 
$[t_2-\Delta t,t_2]$. 
Let $I_1$ and $I_2$  be the integration values  over the former and
the latter region, respectively. By a straightforward calculation,
we can estimate $I_2$ as 
\begin{eqnarray}
I_2& \simeq & 2g A_\lambda'(t_2)(\fc+A_\lambda(t_2))
\ep^{\gamma_2}
\int_{0}^{\ep^{-\gamma_2}\Delta t} ds G_0(s)\Theta(\ep^{\gamma_1}s)
\nonumber \\
& & +g\int_{t_2-\Delta t}^{t_2} ds_2
A_\lambda'(s_2)(A_\lambda^2(t_2-s_2)+2A_\lambda(t_2-s_2) 
A_\lambda(s_2))
\label{I2}
\end{eqnarray}
in the lowest order evaluation. We next combine the second line
of (\ref{I2}) 
with $I_1$ and return it to the original form. 
As the result, we obtain
\begin{eqnarray}
F_\ep(\ep^{-\gamma_2} t_2;\phiu) 
\simeq  
& & A_\lambda (t_2)-g(1-\fc)A_\lambda^2( t_2) 
+g \int_{0}^{t_2}ds_2 {A_\lambda}^2(t_2-s_2) {A_\lambda}'(s_2)
\nonumber \\
& &+2g A_\lambda'(t_2)(\fc+A_\lambda(t_2))
\ep^{\gamma_2}
\int_{0}^{\ep^{-\gamma_2}\Delta t} ds G_0(s)\Theta(\ep^{\gamma_1}s),
\label{Fsecond}
\end{eqnarray}
where higher order terms are ignored. 
With the aid of  (\ref{def_A}), we rewrite the first line of 
(\ref{Fsecond}) as $\ep A(\lambda t_2)/\gc$.
Furthermore, from an estimation
\begin{equation}
\int_{0}^{\ep^{-\gamma_2}\Delta t } ds G_0(s)\Theta(\ep^{\gamma_1}s)
\simeq c_0 \ep^{-\gamma_1(1-a)}\int_0^\infty ds s^{-a}\Theta(s),
\label{est}
\end{equation}
which is valid in the limit $\ep \to 0$, 
the second line of (\ref{Fsecond}) turns out to be 
of $O(\epsilon^{\gamma_2-\gamma_1(1-a)})$. 
These results lead to  (\ref{F_eporder}).
We  also find that the higher order terms we have neglected in 
(\ref{Fsecond}) are of $O(\ep^{\gamma_2-\gamma_1(1-2a)})$ 
by an estimation similar to (\ref{est}).

\section{Derivation of  (\ref{Q:det})}\label{app:eqQ}

We take $\Dt = \ep^{-\alpha'}$, where $\alpha'$ satisfies 
$\alpha' <\gamma_1-1/2$. We also define 
\begin{eqnarray}
{w}(t_1)\equiv c_0t_1^{-a}\Theta_*(t_1)-\lambda t_1.
\end{eqnarray}
Then, for  sufficiently small $\epsilon$, 
$h_\ep(t)\equiv \phi_{\ep}(t) - \fc$  is expressed by
\begin{equation}
h_\ep(t)=G_0(t)+O(\ep)
\label{hex1}
\end{equation}
for $0 \le t \le \Dt$, and 
\begin{equation}
h_\ep(t)=\ep^{1/2}w(\ep^{\gamma_1} t)+\ep^{\alpha}
\bar \varphi_1^{(\alpha)}(\ep^{\gamma_1}t)
\label{hex2}
\end{equation}
for $\Dt \le t \ll \ep^{-\gamma_2}$. 

By substituting $\phi_{\ep}(t)=\fc+h_\ep(t)$ into (\ref{MCT}),
we can write  the equation for $h_\ep(t)$. 
The further substitution of (\ref{hex1}) 
and (\ref{hex2}) into the obtained equation for $h_\ep$ yields 
\begin{eqnarray}
&&
\ep \lr{2 
{w}^2(t) +1 /8 + 4 \int_{\Dt \ep^{\gamma_1}}^{t_1-\Dt\ep^{\gamma_1}} 
ds_1 \lr{ {w}(t_1-s_1)- {w}(t_1)} {w}'(s_1)}
\nm
&&
= O\left(
\ep^{1/2+\gamma_1-\alpha'}, \ep^{3/2},\ep^{\gamma_1},\ep^{\alpha+1/2} 
\right).
\label{eq_tilQ}
\end{eqnarray}
Extracting the $\ep$-independent terms in the limit $\ep \to 0$, 
we obtain 
\begin{eqnarray}
2 {w}^2(t_1) +1/8 + 4 \int_0^{t_1}
ds_1 \lr{ {w}(t_1-s_1)- {w}(t_1)} 
{w}'(s_1)=0.
\end{eqnarray}
We substitute $w(t_1)=Q(t_1)-\lambda t_1$ into this equation.
The result becomes  (\ref{Q:det}).

\section{Short time expansion of  $Q$}\label{app:exp}

We assume the form
\begin{equation}
Q(t_1)=\sum_{k=0}^\infty q_k t_1^{a(2k-1)}+\lambda t_1.
\label{Qexp}
\end{equation}
By substituting (\ref{Qexp}) into (\ref{Q:det}), 
we can determine $q_k$ $(k \ge 1)$ recursively from $q_0=c_0$. 
Concretely, the recursion equation becomes 
\begin{equation}
q_1=-\frac{1}{64 c_0 (V_{0,1}-1/2)}
\end{equation}
and 
\begin{equation}
q_{k+1}=-\frac{1}{2q_0(V_{0,k+1}-1/2)}
\left[\sum_{j=1}^k q_j q_{k+1-j} (V_{j,k+1-j}-1/2 ) \right]
\end{equation}
for $k \ge 1$, where 
\begin{equation}
V_{m,n}=\frac{{p(2m-1)}{p(2n-1)}}{p(2m+2n-2)}
\end{equation}
with $p_n=\Gamma(1+an)$.


\end{document}